\documentstyle[epsfig]{elsart}

\newcommand{\apj}{Astrophys. J.}

\newcommand{\be}{\begin{equation}}
\newcommand{\ee}{\end{equation}}

\newcommand{\nl}{\par\noindent}

\def\simlt{\lower.5ex\hbox{\ltsima}}
\def\gtsima{$\; \buildrel > \over \sim \;$}
\def\simgt{\lower.5ex\hbox{\gtsima}}

\def\simlt{\lower.5ex\hbox{\ltsima}}
\def\gtsima{$\; \buildrel > \over \sim \;$}
\def\simgt{\lower.5ex\hbox{\gtsima}}

\def\cm{{\rm\,cm}}

\def\ergcm2{\ {\rm erg~cm^{-2} }}
\def\ergscm2{\ {\rm erg~s^{-1}~cm^{-2} }}

\def\cm2s{\ cm^2 ~s^{-1} }

\def\s{\ifmmode \widetilde \else \~\fi}
\def\={\overline}

\def\spose#1{\hbox to 0pt{#1\hss}}

\def\lta{\mathrel{\spose{\lower 3pt\hbox{$\mathchar"218$}}
     \raise 2.0pt\hbox{$\mathchar"13C$}}}
\def\gta{\mathrel{\spose{\lower 3pt\hbox{$\mathchar"218$}}
     \raise 2.0pt\hbox{$\mathchar"13E$}}}
\def\mincir{\ \raise -2.truept\hbox{\rlap{\hbox{$\sim$}}\raise5.truept  
\hbox{$<$}\ }}                                                          %
\def\magcir{\ \raise -2.truept\hbox{\rlap{\hbox{$\sim$}}\raise5.truept  %
\hbox{$>$}\ }}                                                          %
\def\simlt{\ \raise -2.truept\hbox{\rlap{\hbox{$\sim$}}\raise5.truept   
\hbox{$<$}\ }}                                                          %
\def\simgt{\ \raise -2.truept\hbox{\rlap{\hbox{$\sim$}}\raise5.truept   %
\hbox{$>$}\ }}                                                          %
\def\newline{\par\noindent}

\def\s-z{S-Z}

\begin{document}
\begin{frontmatter}
\title{Cosmic Rays, Radio Halos and Non-Thermal X-ray Emission
in Clusters of Galaxies}

\author[Chicago]{Pasquale Blasi} and
\author[Roma]{Sergio Colafrancesco}
\address[Chicago]{Dept. of Astronomy \& Astrophysics, E. Fermi Inst.,
University of Chicago, Chicago, IL 60637-1433, USA}
\address[Roma]{Osservatorio Astronomico di Roma,
Via dell'Osservatorio 2, I-00040 Monteporzio, Italy}

\begin{abstract}

We calculate the flux of radio, hard X-ray and UV radiation
from clusters
of galaxies as produced by synchrotron emission and Inverse
Compton Scattering of electrons generated as secondaries
in cosmic ray interactions in the intracluster medium.
Both the spatial distribution of cosmic rays due to their
diffusion and the spatial distribution of the intracluster gas are
taken into account.

Our calculations are specifically applied to the
case of the Coma cluster.
The fluxes and spectra of the radio halo emission and of the hard
X-ray excess from Coma can be explained in
this model if an average magnetic field
$B \sim 0.1\mu G$ is assumed.
However, such a low value for the intracluster magnetic field
implies a large
cosmic ray energy density which in turn is responsible,
through neutral pion decay, for
a gamma ray flux above $100$ MeV which exceeds
the EGRET upper limit.
This gamma ray bound can be relaxed if the hard X-ray excess
and the radio halo emission from Coma are not due to the same 
population of electrons. 

We finally stress the unique role that the
new generation gamma ray satellites will play to
discriminate among different models for the non thermal
emission in clusters of galaxies.

\end{abstract}
\end{frontmatter}

\section{Introduction\label{intro}}

The origin of the non--thermal, diffuse emission (radio, X-ray and UV) in
galaxy clusters still remains poorly understood.
The general consensus is that the radio and X-ray non--thermal
emission are
likely to be produced by a population of relativistic electrons
via synchrotron emission
and Inverse Compton Scattering (hereafter ICS)
off the Microwave Background photons, respectively.
However, the origin, acceleration and propagation of these
relativistic electrons are still subject to debate.
If electrons are accelerated in discrete sources
in the central regions of clusters, the severe energy losses and the
diffusive motion force the electrons to cover at most distances
of order of a few kpc from the source, while diffuse radio emission
is clearly observed on Mpc scales \cite{fg97}.
To solve this problem, reacceleration models have been invoked
(see \cite{gfer} for a discussion),
though the nature of the {\it in situ} acceleration is not yet
completely clear.

A natural solution to the problem mentioned above was first proposed
in \cite{dennison} and \cite{vestrand}, where the electrons
responsible for
the non-thermal emission were produced {\it in situ} as
secondary products of cosmic
ray nucleons interacting with the protons in the intracluster medium
(hereafter ICM).
Cosmic rays (hereafter CR)
in clusters of galaxies are, in fact,
practically free from energy losses and can
reach large distances from the source.
An interesting consequence of this model is that not only electrons but
also gamma rays (through the decay of neutral pions) and neutrinos are
also produced in the same interactions \cite{dar,bbp,cb}.
We show here that this point can
strongly constrain the validity of the secondary electron model
(hereafter SEM) as the
explanation of the diffuse, non-thermal emission from galaxy clusters.
The SEM in his original version \cite{dennison,vestrand}
was recently questioned \cite{cinesi} as the
explanation of the diffuse radio halo of Coma, because of the too steep
CR spectrum [$Q(E)\sim E^{-(2.5-2.7)}$] required to fit the data.

The aim of this paper is to calculate the flux and the spectra of
the non-thermal radio, UV  and
hard X-ray emission from Coma in the framework of the SEM,
addressing the following points, not considered in previous works:

{\it i})  the calculations in \cite{dennison,vestrand,cinesi} are
all carried out in the assumption that CR are permanently confined
in the cluster volume. Indeed, as shown in \cite{bbp} and \cite{cb}
this is true only for CR with energy below a {\it confinement energy}
$E_c$ (see Section 2 below),
which depends quite strongly on the diffusion
coefficient in the ICM;

{\it ii}) in previous calculations \cite{dennison,vestrand,cinesi},
a spatially homogeneous equilibrium distribution of CR was always
assumed. However, if CR are mainly contributed by a small number of
discrete powerful
sources in the cluster core (as it seems to be the case in Coma),
the distribution of CR due to diffusion is strongly inhomogeneous
(for instance in the case of a single source the CR distribution goes
as $1/r$ if $r$ is the distance to the source).
Similar conclusions hold in case of shocks produced by (sub)cluster
merging or during the cluster collapse, as sources of CR in the
clusters.
We consider here both the cases of a single source and of a
spatially homogeneous
injection of CR in the ICM;

{\it iii}) the observable fluxes of non-thermal radiation are calculated as a
convolution
of the distribution of CR and of the targets for CR interactions,
which are provided by the IC gas protons.
Therefore the spatial distribution
of the IC gas, never considered in previous calculations,
can have a relevant role and needs to be considered;

{\it iv}) in the previous works \cite{dennison,vestrand,cinesi}
a quite steep spectrum,
$J(\nu) \sim \nu^{-1.34}$, for the Coma radio halo was considered for
the comparison with the model predictions.
This was the main reason to call for a correspondingly steep CR spectrum
and eventually claim to rule out the SEM \cite{cinesi}.
However recent observations (see, e.g.,  \cite{fg97})
show that the radio spectrum from Coma is
well fitted by a power law with power index $\sim 1.16$ up to a
frequency $1.4$ GHz, while there is a still controversial evidence for
a steepening at higher frequencies (see \cite{deiss} \cite{bowyer}; see also
\cite{fg97} for a review).

{\it v}) we use here a detailed treatment for the $pp$ collisions,
which is particularly important in the calculation of the fluxes
of gamma and UV radiation from clusters of galaxies.

In addition to the points listed above, there are some recent 
data that introduce new limits on the SEM.
In particular, the combination
of the EGRET limit on the gamma ray emission from Coma
\cite{sreek} and the recent
detection of an hard X-ray non-thermal tail above $20$ keV from Coma,
obtained by the SAX \cite{fusco98} and the
RXTE \cite{rg98} experiments,
allows us to strongly constrain the validity of the SEM as a
combined explanation of both the radio halo emission  
and the hard X-ray excess. 
Furthermore, we also consider the still debated evidence of a 
diffuse UV emission from
Coma \cite{lieu96,bowyer} as a possible indication of ICS emission from
very low energy electrons.

The structure of the paper is the following:
in Section 2 we discuss the CR propagation in clusters of galaxies;
in Section 3 we derive
the secondary electron spectrum from CR collisions in the ICM.
We present our calculations for the non-thermal,
diffuse radio and X-ray emission in Section 4.1 and 4.2, respectively.
In Section \ref{gamma} we also derive the expected gamma-ray flux produced
by the same CR interactions responsible for the secondary electrons.
In Section \ref{coma} the calculations are applied and discussed in
the case of Coma. Our conclusions are presented in
the final Section \ref{discussion}.
Throughout the paper we use an dimensionless Hubble parameter
$h=(H_0/100 ~km ~s^{-1} ~Mpc^{-1}) =0.6$ and
$\Omega_0=1$, unless otherwise specified.

\section{Cosmic ray propagation in the intracluster medium \label{prop}}

Following \cite{bbp,cb}, we first assume that the accelerated CRs in
clusters are mainly supplied by radio galaxies -- or more generally
by active galaxies -- and/or by the possible shocks occurring
predominantly in the central regions of galaxy clusters.
As estimated in \cite{bbp,cb} we can expect nearly one of these active
galaxies per rich cluster, on average, and also we can expect that
there is at least one major shock after its formation \cite{quilis}
in the central regions of the clusters.
Actually, this seems to be the situation in the
Coma cluster (see \cite{briel}).
For simplicity, we consider here a CR source
in the central region of the cluster irrespective of its peculiar nature.
We will also discuss the more general case of an extended
distribution of cluster CR sources in a forthcoming paper \cite{bc98}.
In the end of this section we consider the case of a spatially
homogeneous injection of CR as an extreme case opposite to that
of a single source.

In the typical magnetic fields present in clusters of galaxies
(see \cite{kimkro} and \cite{kro} for reviews), the propagation
of CR is diffusive and can be described by the transport equation:
\begin{equation}
\frac{\partial n_p(E_p,r,t)}{\partial t} - D(E) \nabla^2 n_p(E_p,r,t) -
\frac{\partial}{\partial E_p}
\left[ b(E_p) n_p(E_p,r,t)\right] = Q(E_p)\delta (\vec {r}),
\label{eq:transport}
\end{equation}
\noindent
where $n_p(E_p,r,t)$ is the density of CRs with energy $E_p$
at distance $r$ from the source, $Q_p(E_p)=Q_0 p_p^{-\gamma}$
[with $p_p=(E_p^2-m_p^2)^{1/2}$ being the proton momentum]
is the assumed spectrum of the CR source and $b(E_p)$ is the
rate of energy losses. 
The normalization constant, $Q_0$, is
related to the CR luminosity, $L_p$, by
$$
Q_0\int_0^{E_p^{max}} dT_p ~T_p p_p^{-\gamma}=L_p~,
$$
where $T_p=E_p-m_p$ is the kinetic energy of the proton.
The diffusion coefficient, $D(E_p)$, has been explicitely assumed to be
independent of $r$ because we consider an average magnetic field
uniformly distributed in the ICM.
As for the diffusion coefficient,
we adopted the same approach as in \cite{cb}.
Specifically, we assume that the
spectrum of the fluctuations in the IC magnetic field
is described by a Kolmogorov spectrum, $P(k)=P_0 (k/k_0)^{-5/3}$,
and we use the following expression to relate the
diffusion coefficient $D(E_p)$ to $P(k)$:
$$
D(E_p)=\frac{1}{3} r_L c \frac{B^2}{\int_{1/r_L}^\infty dk P(k)}
$$
\begin{equation}
=2.3\times 10^{29} E_p(GeV)^{1/3} B_\mu^{-1/3} cm^2/s
\left(\frac{l_c}{20kpc}\right)^{2/3} ~.
\label{eq:diff}
\end{equation}
Here $r_L$ is the Larmor radius of the particles with energy
$E_p$ and $B$ is the rms magnetic field strength
($B_\mu$ is its value expressed in $\mu G$).
The normalization constant $P_0$ is obtained
requiring that the total energy density in the field equals $B^2$.
Here $l_c\approx 1/k_0$ is the size of the largest eddy in the
magnetic field.

In clusters of galaxies eq. (\ref{eq:transport}) simplifies appreciably
because the term of energy losses can be neglected. In fact, for a typical
rich cluster with IC gas mass, $M_{gas}\simeq 10^{14}~M_{\odot}$, and
total mass, $M_{tot} \sim 10^{15} ~M_{\odot}$,
over a distance comparable with the Abell radius $R_A= 1.5/h$
Mpc, the average IC gas density turns out to be
${\bar n}_H\sim 3 \cdot 10^{-4}~cm^{-3} ~h^2$,
which corresponds to a column density of
$\sim 6~g/cm^2\ll x_{nucl}=(m_p/\sigma_{pp})\approx 50 \div 100~g/cm^2$
(here $\sigma_{pp}\approx 3\times 10^{-26} cm^2$ is the cross section
for $pp$ collisions and $m_p$ is the proton mass).
This estimate is found assuming CR confined in the cluster for times
comparable with the cluster age. For non confined CR, the IC gas
column density is even lower.

In this framework, the solution of eq.(\ref{eq:transport}) found at 
the present
time, $t_0$, after the source started to produce CR is given by:
\begin{equation}
n_p(E_p,r)=\frac{Q_p(E_p)}{D(E_p)}\frac{1}{2\pi^{3/2} r}
\int_{r/r_{max(E_p)}}^{\infty} dy e^{-y^2} ~,
\label{eq:sol}
\end{equation}
where the source term has been assumed active for all the age, $t_{cl}$
of the cluster (which is comparable, for our purposes,
to  the age, $t_0$, of the universe).

We introduce the quantity $r_{max}(E_p)=\sqrt{4D(E_p)t_0}$ which represents
the average maximum distance that a particle with energy
$E_p$ can diffuse away from the source in the time $t_0$.
If $r\ll r_{max}(E_p)$, then the lower limit in the
integral in eq.(\ref{eq:sol}) above is small and
eq.(\ref{eq:sol}) gives the well known stationary solution:
\begin{equation}
n_p(E_p,r)=\frac{Q_p(E_p)}{D(E_p)}\frac{1}{4\pi r}.
\label{eq:solapp}
\end{equation}
At distances larger than $r_{max}(E_p)$ the density of particles with
energy $\leq E_p$ is suppressed.

If $R_{H}$ is the confinement size of a cluster
(here we consider $R_H \sim $ the radio halo size), then the solution of the
equation $r_{max}(E_c)=R_{H}$ gives an estimate of the maximum energy,
$E_c \sim (R_H^2/4 D_0 t_0)^{1/\eta}$, for which CRs
are confined inside the cluster.
Thus,  particles with energy lower than $E_c$
need times larger than $t_0$ to leave the cluster,  and
particles with energy larger than $E_c$ diffuse away from the cluster
in the time $\sim R^2_{H}/4D(E_p)$.
In other words, clusters of
galaxies can be considered as {\it closed boxes} only for CRs
having energy $E_p \simlt E_c$ \cite{bbp,cb}.

Finally, we discuss here also the extreme case of a homogeneous injection of CR in the
ICM. In this case the equilibrium CR distribution is simply given by
\begin{equation}
n_p(E_p)=K\frac{\epsilon_{tot}}{V}p_p^{-\gamma}
\label{eq:sol_homo}
\end{equation}
where again the constant $K$ is obtained requiring that
$$
K\int_0^{E_p^{max}} dT_p T_p p_p^{-\gamma}=\epsilon_{tot}~,
$$ 
and $\epsilon_{tot}$ is the total energy injected in the cluster
volume $V$ in the form of CR.
This neglects the leaking of CR
from the boundary of
the cluster, which is equivalent to assume that the relevant
CR are confined in the cluster. 
This assumption is justified, because even if
CR with high energy do escape, they do not appreciably affect the
energy balance in the cluster for injection spectra steeper than
$E_p^{-2}$ at high energy. 
Note that eq. (\ref{eq:sol_homo}) is 
no longer valid close to the cluster boundary.

\section{Secondary electrons in pp collisions\label{second}}

CRs interact with the IC gas protons and produce electrons
($e^+e^-$), neutrinos and gamma rays through the decays of charged and
neutral pions, respectively (see, e.g., \cite{bbp,cb}).
In this section we concentrate on
the electron component which is relevant for the calculations
of the radio, UV  and hard X-ray emission.

The spectrum of secondary electrons with energy $E_e$ at distance $r$
from the CR source is readily calculated by convolution of the
proton, pion and muon spectra and is given by:
$$
q_e(E_e,r)= \frac{m_{\pi}^2}{m_{\pi}^2-m_{\mu}^2} n_H(r) c \cdot
$$
\begin{equation}
\int_{E_e}^{E_p^{max}} dE_\mu \int_{{E_\pi}^{min}}^{{E_\pi}^{max}}
dE_\pi \int_{E_{th}(E_\pi)}^{E_p^{max}} dE_p~F_\pi(E_\pi,E_p)
F_e(E_e,E_\mu,E_\pi) n_p(E_p,r) ~,
\label{eq:source}
\end{equation}
where $E_p^{max}$ is the maximum proton energy (note that 
our calculations are insensitive to the exact value of $E_p^{max}$),
$E_{th}(E_\pi)$ is the threshold energy for the production of pions with
energy $E_\pi$ and we put
$$
E_\pi^{min}=\frac{2 E_\mu}{(1+\beta)+\xi (1-\beta)}~;~~~
E_\pi^{max}=min\left\{E_p^{max},
\frac{2 E_\mu}{(1-\beta)+\xi (1+\beta)}
\right\}~,
$$
where $\xi=m_\pi^2/m_\mu^2$ and $\beta$ is the velocity of muons in units 
of the light speed.
The quantity $n_p(E_p,r)$ is the equilibrium CR spectrum at distance
$r$ from the source, as given by eq. (\ref{eq:sol})
for a single CR source or from eq. (\ref{eq:sol_homo}) in the
case of a homogeneous CR injection.
The quantity $n_H(r)$ is the density profile of the IC gas
as a function of the radial distance from the cluster center and is
well fitted by a King profile,
$n_H(r) = n_H^0 \left[ 1+\left( \frac{r}{r_0} \right)^2
\right]^{-3\beta_{IC}/2},$
where $n_H^0$ is the central IC gas density, $r_0$ is the 
core radius and $\beta_{IC}$ is a phenomenological parameter
in the range $0.7 \simlt  \beta_{IC} \simlt 1.1$
(see \cite{sarazin} for a review).\par
The physics of the interaction is contained in the functions
$F_\pi(E_\pi,E_p)$ (the spectrum of pions produced in a CR interaction
at energy $E_p$ in the laboratory frame) and $F_e(E_e,E_\mu,E_\pi)$
(the spectrum of electrons from the decay of a muon of energy $E_\mu$
produced in the decay of a pion with energy $E_\pi$). The electron
spectrum is given by the following expression:
$$
F_e(E_e,E_\mu,E_\pi)=\frac{1}{\beta E_\mu}\times
$$
\begin{equation}
\times 
\left\{
        \begin{array}{l l}
                2\left(\frac{5}{6}-\frac{3}{2}\lambda^2+\frac{2}{3}
 \lambda^3\right)-P_\mu\frac{2}{\beta}\left[\frac{1}{6}-
\left(\beta+\frac{1}{2}\right)\lambda^2+\left(\beta+\frac{1}{3}\right)
\lambda^3\right]~~~~~ &
              \mbox{if $\frac{1-\beta}{1+\beta}\leq \lambda\leq 1$},\\

       \frac{4\lambda^2\beta}{(1-\beta)^2}\left[3-\frac{2}{3}\lambda
	\left(\frac{3+\beta^2}{1-\beta}\right)\right]-\\
\frac{4P_\mu}{1-\beta}\left\{\lambda^2(1+\beta)-\left[
\frac{1}{2}+\lambda (1+\beta)\right] \frac{2\lambda^2}{1-\beta}+
\frac{2\lambda^3 (\beta^2+3)}{3(1-\beta)^2}\right\} &
                     \mbox{if $0\leq\lambda\leq \frac{1-\beta}{1+\beta}$},
        \end{array}
        \right\}
\end{equation}
where we put
\begin{equation}
P_\mu=P_\mu(E_\pi,E_\mu)=\frac{1}{\beta}\left[
\frac{2E_\pi \xi}{E_\mu(1-\xi)}-\frac{1+\xi}{1-\xi}\right],
\end{equation}
and $\lambda=E_e/E_\mu$.
The above expression for $F_e$ takes into account that the
muons produced from the decay of pions are fully polarized (this
is the reason why the pion energy $E_\pi$ appears in the expression for
the electron spectrum from the muon decay).

Determining the pion distribution is not trivial
in particular in the low energy region (pion energies close to the
mass of the pions) where not many data are available. A satisfactory
approach to the low energy pion production was proposed in
\cite{dermer} and recently reviewed in \cite{strong} in the context
of the {\it isobaric model}. The detailed and lengthy expressions
for $F_\pi$ that we used are reported and discussed in details in
\cite{strong} (see their Appendix). Thus, following \cite{strong}, 
we use here their
model for collisions at $E_p\lta 3$ GeV. For $E_p\gta 7$ GeV we use
the scaling approximation which can be formalized writing the
differential cross section for pion production as
\begin{equation}
d \sigma/ d E_{\pi} = (\sigma_0/E_{\pi}) f_{\pi^\pm}(x),
\label{eq:scal}
\end{equation}
where $\sigma_0=3.2\cdot 10^{-26}~cm^2$, $x=E_{\pi}/E_p$.
The scaling function $f_{\pi^\pm}(x)$ is given by
\begin{equation}
f_{\pi^\pm}(x)=1.34 (1-x)^{3.5} + e^{-18x}.
\label{eq:fpi}
\end{equation}
In this case the function $F_\pi$ coincides with the definition of
differential cross section given in eq. (\ref{eq:scal}).

The electron spectrum rapidly reaches its equilibrium
configuration mainly due to synchrotron and ICS energy
losses at energies larger than $\sim 150$ MeV and mainly due
to Coulomb energy losses at smaller energies. The equilibrium
spectrum can be calculated solving the transport equation
\begin{equation}
-\frac{\partial}{\partial E_e} \left[n_e(E_e,r)b(E_e)\right]
= q_e(E_e,r)
\end{equation}
where 
$n_e(E_e,r)$ is the equilibrium electron distribution and
$$
b_e(E_e) = \left(\frac{dE_e}{dt}\right)_{ICS} +
\left(\frac{dE_e}{dt}\right)_{syn} +
\left(\frac{dE_e}{dt}\right)_{Coul}
= b_0(B_{\mu}) E_e^2 + b_{Coul}
$$
is the rate of energy
losses per unit time at energy $E_e$.
Here, we put
$b_0(B_{\mu})= (2.5\cdot 10^{-17} + 2.54 \cdot 10^{-18} B_{\mu}^2)$
and $b_{Coul}= 7\times 10^{-16} n_H(r)$
(if $b_e$ is given in units of GeV/s). In the expression for
$b_{Coul}$, the IC gas density $n_H(r)$ is given in units of $cm^{-3}$.

\section{Non thermal emission from galaxy clusters}

In this section we derive the non-thermal radio,
hard X-ray and gamma ray spectra of galaxy clusters as 
obtained in the context of
the SEM for the case of a single CR source.

\subsection{The diffuse radio emission}\label{radio}

The calculation of the radio emissivity per unit volume is
performed here in the simplified assumption that electrons with energy $E_e$
radiate at a fixed frequency given by \cite{longair}:
\begin{equation}
\nu \approx 3.7\cdot 10^6 B_{\mu} E_e^2(GeV) Hz.
\label{eq:freq}
\end{equation}
This approximation introduces negligible errors in the final result and
has the advantage of making it of immediate physical interpretation.\par
The radio emissivity at frequency $\nu$ and at distance $r$ from
the cluster center can be calculated as
\begin{equation}
j(\nu,r)  = n_e(E_e,r) \left(\frac{dE_e}{dt}\right)_{syn}~
\frac{dE_e}{d\nu}.
\label{eq:j2}
\end{equation}

The observed flux of radio emission from the cluster is
obtained by integration of $j(\nu,r)$ over the cluster
volume.
Though the numerical calculations have been carried
out using the exact solution of eq.(\ref{eq:sol})
for the CR equilibrium spectrum,
some useful information can be extracted
from the following  simplified picture.
Let us assume that at fixed proton
energy, $E_p$, particles with energy larger
than $E_p$ are the only ones able to diffuse out to
distances $r>r_{max}(E_p)$ [at first approximation
this assumption is correct just by the definition of $r_{max}(E_p)$].
Let us also assume for simplicity a constant IC gas density,
$\bar n_H$. Most of the electrons responsible for the
radio emission are produced in CR interaction at energies
in the scaling regime [eqs. (\ref{eq:scal}-\ref{eq:fpi})].
Within these assumptions it is straightforward to show that
$j(\nu,r)\propto (1/r) \nu^{-(\gamma+\eta)/2}$ and that the
integrated diffuse radio flux reads:
\begin{equation}
J(\nu)\propto \nu^{-(\gamma+\eta)/2} \int_0^{r_{max}(\nu)}
dr ~r.
\label{eq:rough}
\end{equation}
The electron energy and the radio frequency are related
through eq. (\ref{eq:freq}) and in this simplified approach we
can assume that protons with energy $E_p\sim 10E_e$
produce electrons with energy $E_e$. This means that
the distance $r_{max}(\nu)$ in eq. (\ref{eq:rough})
is proportional to $\nu^{\eta/4}$ for collisions involving
CR confined in the cluster, while it is simply equal to the
confinement radius $R_H$ for non confined CR.
In the first case, the observed spectrum is $J(\nu) \propto
\nu^{-\gamma/2}$,
independent on the details of the diffusion (this is the
same result highlighted in \cite{bbp} and \cite{cb} for
gamma rays) while, in the other case, the spectrum is steepened as
$J(\nu) \propto \nu^{-(\gamma+\eta)/2}$ above a frequency
$\nu^*$ which is dependent on the diffusion coefficient and on
the spatial scale of the confinement region.
The situation is indeed more complicated in
any realistic case, but the main features and in
particular the high frequency steepening are still present. 
In particular, the introduction of a realistic
King profile for the IC gas density distribution in the cluster
translates into a decrease of the effective frequency $\nu^*$
with respect to the case of a uniform IC gas distribution.

\subsection{The diffuse, non-thermal X-ray and UV emissions \label{hard}}

The relativistic electrons responsible for the radio halo emission
also emit X-rays and UV photons through ICS off the Microwave Background
photons. As in the case of the synchrotron emission, also for ICS
we adopt the approximation that electrons radiate at a single energy,
given by
\begin{equation}
E_X=2.7~keV ~E^2_e(GeV).
\label{eq:ex}
\end{equation}
Electrons with energy in excess of a few GeV radiate in the
hard X-ray range, while electrons with energy smaller than $\sim400$
MeV produce soft X-rays and UV photons.\par
The non-thermal X-ray/UV emissivity at distance $r$ from the
CR source is
\begin{equation}
\phi_X(E_X,r) = n_e(E_e,r) \left(\frac{dE_e}{dt}\right)_{ICS}~
\frac{dE_e}{dE_X} ~.
\label{eq:phi2}
\end{equation}
\noindent
In complete analogy with the case of the radio emission,
the integrated non-thermal X-ray flux, $\Phi_X(E_X)$ is
\begin{equation}
\Phi(E_X,r) =
\int_0^{R_H} dr ~4 \pi r^2 \phi_X(E_X,r) ~.
\label{eq:phi2_1}
\end{equation}
Considerations similar to those presented in the previous section hold
for the steepening of the hard X-ray spectrum from ICS.

\subsection{The gamma ray emission}\label{gamma}

Gamma ray emission from clusters of galaxies 
is mainly produced by the decay of neutral pions
(see, e.g., \cite{bbp,cb}) and by bremsstrahlung emission
of secondary electrons (note that also primary electrons 
contribute to the bremsstrahlung flux, so that the contribution
calculated here is a lower limit).

The emissivity in gamma rays at distance $r$ and
energy $E_\gamma$ is given by
\begin{equation}
j_\gamma^{\pi^0}(E_\gamma,r)=2 n_H(r) c
\int_{E_\pi^{min}(E_\gamma)}^{E_p^{max}}
dE_\pi \int_{E_{th}(E_\pi)}^{E_p^{max}}
dE_p F_{\pi^0}(E_\pi,E_p)\frac{n_p(E_p,r)}{(E_\pi^2+m_\pi^2)^{1/2}},
\label{eq:gamma1}
\end{equation}
where $E_\pi^{min}(E_\gamma)=E_\gamma+m_{\pi^0}^2/(4E_\gamma)$.
We refer to \cite{strong} for the expression of $F_{\pi^0}(E_\pi,E_p)$
in the low energy collisions ($E\leq 3$ GeV), while we use 
the scaling approach
given in eqs. (\ref{eq:scal}) and (\ref{eq:fpi}) for $E_p>7$ GeV,
with $f_{\pi^0}=(1/2)f_{\pi^\pm}$.\par
The flux of gamma rays due to bremsstrahlung of secondary electrons
is given by
\begin{equation}
j_\gamma^{brem}(E_\gamma,r)=n_H(r) c\int_{E_\gamma}^{E_p^{max}}
dE_e n_e(E_e,r) \frac{d\sigma}{dE_\gamma}(E_e,E_\gamma),
\end{equation}
where the differential cross section can be written as
\begin{equation}
\frac{d\sigma}{dE_\gamma}(E_e,E_\gamma)=2.6\cdot 10^{-26}
\frac{1}{E_e}\Lambda(v)~~~ cm^2 GeV^{-1},
\end{equation}
and the quantity $\Lambda(v)$ is:
\begin{equation}
\Lambda(v)=v+\frac{1-v}{v}\left(\frac{4}{3}+2 b\right)
\end{equation}
with $b\approx 0.01$.

The total amount of gamma ray emission at energy $E_\gamma$
from a single cluster is obtained, as usual, by volume integration:
\begin{equation}
J_\gamma(E_\gamma)=\int_0^{R_H} dr~ 4\pi r^2 j_\gamma(E_\gamma,r),
\label{eq:gamma2}
\end{equation}
where $j_\gamma(E_\gamma,r)= j_\gamma^{\pi^0}(E_\gamma,r)+
j_\gamma^{brem}(E_\gamma,r)$.

\section{The case of the Coma cluster} \label{coma}

As described in the previous sections the SEM implies that:
\nl
{\it i)} a flux of radio emission is produced due to
the synchrotron emission of electrons produced in CR
interactions;
\nl
{\it ii)} secondary electrons with $E\geq 1$ GeV radiate in the
X-ray energy range due to ICS on the photons of the
microwave backgound radiation;
\nl
{\it iii)} secondary electrons with $E\leq 400$ MeV produce
UV photons in the energy range $E_{UV}\leq 0.4$ keV.

{\it iv)} gamma rays are copiously
produced in the decays of the neutral pions.
\nl
The last point {\it iv)} represents, at the same time,
a unique signature of
the SEM and a way to impose strong constraints on this model.

In this section we analyze the predictions of the SEM
for the Coma cluster and we compare them with the available data in the
radio \cite{gfer}, hard X-ray \cite{fusco98} \cite{rg98},
UV \cite{lieu96,bowyer,lieu98} bands and with the EGRET upper limit on the 
gamma ray emission at $E>100$ MeV \cite{sreek}.

Let us begin with the case of a single source of CR.
As we stressed in Section 2, the CR diffusion coefficient, $D(E_p)$,
is uncertain and there are not yet reliable constraints on
its functional dependence on the CR energy.
We adopt here
a Kolmogorov spectrum of the fluctuations in the IC
magnetic field and derive an explicit form for the
diffusion coefficient, as given in eq. (\ref{eq:diff}).
This expression has the advantage to relate the
diffusion coefficient $D(E_p)$ directly to observable quantities
like the coherence length and the strength
of the magnetic field, while typically these quantities are all
considered as independent parameters.
Following \cite{jaffe}, we fix here $l_c=20$ kpc,
to have a conservative result. The conclusions discussed below will
be strengthened for a larger value of $l_c$.
As for the parameters of the IC gas density profile of Coma, we used
$n_H^0=3\cdot 10^{-3}cm^{-3}$, $r_0=400$ kpc and $\beta_{IC}=0.75$.

The diffuse radio flux for Coma  has been evaluated
according to eq. (\ref{eq:j2}).
We use the CR luminosity, $L_p$,
as a normalization parameter and we fitted it to the observations.
Once the normalization is fixed, the
flux of non-thermal X-rays and gamma rays is easily
calculated from eqs. (\ref{eq:phi2}) and
(\ref{eq:gamma1}-\ref{eq:gamma2}).
This procedure has been repeated for $\gamma=2.1$ and
$\gamma=2.4$, which can be  considered as a lower and an upper
limit to
the power index of the injection CR spectrum, and for three
values of the magnetic field, $B = 0.1$, $1$ and $2$ $\mu G$.

The results of our calculations for the spectrum of the radio halo of
Coma are shown in Fig.~1a ($B_\mu=0.1$),
Fig.~1b ($B_\mu=1$) and Fig.~1c ($B_\mu=2$).
The solid lines refer to $\gamma=2.1$ and the dashed lines to
$\gamma=2.4$.
The data points are taken from \cite{gfer}.
The CR luminosities needed
to fit the radio halo data are reported in Table 1.
It is evident that for large values of the IC magnetic field
($B \sim 1-2\mu G$)
the CR luminosities required to fit the data,
$L_p \sim (0.1-1) \cdot 10^{44}$ erg/s (see Table~1),
are of the order of the
typical CR powers predicted in clusters of galaxies (see
\cite{bbp,cb}).
For low values of $B \sim 0.1 - 0.2 ~\mu G$, considerably
larger CR luminosities, $L_p \simgt 5 \cdot 10^{45}$ erg/s,
are required by the observations.

The allowed range of parameters is drastically reduced
when the analysis is not limited to the radio halo data alone but
include also the recent detection of hard X-ray tails
and the existing EGRET upper limit on the
gamma ray emission from Coma.
Our predictions for the non-thermal X-ray emission from Coma
are plotted in Fig.~2a ($B_\mu=0.1$),
Fig.~2b ($B_\mu=1$) and Fig.~2c ($B_\mu=2$), where the
solid and dashed lines are again for $\gamma=2.1$ and
$\gamma=2.4$, respectively.
The shaded area shows the thermal
bremsstrahlung continuum that fits the HEAO1-A4 and GINGA
data (open triangles) with a temperature of
$T=8.21\pm 0.20$ keV.
The OSSE upper limits are indicated with
open circles and the SAX data points are represented by the
filled squares. The plotted errorbars refer to $90\%$
confidence level.

From Fig.~2 it is clear that only models with small values of
$B_{\mu}$ ($B_\mu \approx 0.1$) can reproduce the observed X-ray flux in
the energy region between $20$ keV and $100$ keV, 
while the contribution of the secondary electrons to the X-ray
flux for larger values of $B_{\mu}$ is negligible.
The interpretation of this result is straightforward:
in fact, contrary to
the diffuse radio flux, the X-ray flux is very weakly dependent on
the value of the magnetic field [see eq.(13)].
Specifically, it depends on $B_\mu$ mainly through the normalization
of the CR spectrum that is required to fit the radio halo data. 
Therefore, low values
of $B_\mu$ imply large CR luminosities $L_p$ (because of the low
radio emissivity) and, in turn, large X-ray fluxes.
On the other hand, for large values of $B_\mu$ electrons radiate
more efficiently and low values of $L_p$ are needed, which in turn
imply low X-ray fluxes.

Recently, also UV emission was detected from Coma \cite{lieu96}, 
and interpreted as 
ICS of electrons with energy between $150$ MeV and $400$ MeV 
\cite{bowyer,lieu98}. 
In the SEM the flux of UV emission due to ICS crucially depends on the
CR injection spectrum below $\sim 1$ GeV, which is poorly known
and may reflect specific conditions in the acceleration region. 
The spectrum of secondary electrons in the energy range 
of interest suffers 
a substantial flattening due to the peculiar shape of the pion 
spectrum at low energy and due to Coulomb energy losses. 
In the assumption, used in this paper, of a CR spectrum which is a 
power law in momentum, 
our prediction of the UV flux for $B_\mu\sim 0.1$ falls short of the
observed flux \cite{lieu96} by a factor $1.5-2$. 

Values $B_{\mu} \sim 0.1$ in the Coma ICM are consistent
with the findings of \cite{fusco98}.
However, in the SEM this result has further implications:
in cases where the X-ray emission becomes
appreciable ({\it i.e.} for low values of $B_\mu$)
with a large $L_p$ needed, the
gamma ray emission predicted for Coma on the basis of eqs.(16-19)
grows linearly with $L_p$ and the gamma ray flux very easily 
exceeds the
EGRET limit as shown in Table 1, where the fourth column
represents the ratio of the flux predicted in the SEM
to the EGRET upper limit from Coma, 
$F_{\gamma}^{EGRET}(E>100MeV) \approx 4
\cdot 10^{-8} ~photons ~s^{-1}~ cm^{-2}$. 
The contribution of secondary electron bremsstrahlung to the gamma
ray emission of Coma is always negligible with respect to the
contribution of gamma rays produced by neutral pion decay (we
report the ratio of the two contributions in the last
column of Table 1).

Our predictions for the differential gamma ray spectra from
Coma are plotted in Fig. 3a ($B_\mu=0.1$), 3b ($B_\mu=1$) and
3c ($B_\mu=2$). The thick lines are the contributions of $\pi^0$
decay while the thin lines represent the bremsstrahlung gamma
rays from the secondary electrons. The solid and dashed lines
are, as usual, for $\gamma=2.1$ and $\gamma=2.4$ respectevely.

How much does this conclusion depend upon the assumption of
a single source of CR? In order to answer this question we
evaluate
the fluxes of radio and hard X-rays for a homogeneous injection
of CR over the cluster volume.
If the observed radio halo spectrum at $\nu \leq 1.4 $ GHz
is taken as a power law with power index $\alpha_r=1.16$ \cite{deiss}, then
the best fit to the data is obtained for $\gamma=2\alpha_r=2.32$.
We use eq. (\ref{eq:sol_homo}) for the equilibrium proton
distribution in the ICM and the results of Section
\ref{second} to determine the relative spectra of
secondary electrons. Radio, X-ray and gamma ray fluxes are calculated
as usual.

The comparison with the radio \cite{gfer}
and hard X-ray \cite{fusco98} data
yields $\epsilon_{tot} \approx
8 \cdot 10^{63}$ erg and $B_\mu\approx 0.1$. The
value of $\epsilon_{tot}$ corresponds, if averaged over the
age of the cluster, to a huge CR luminosity of
$L_p\sim 2\cdot 10^{46}$ erg/s.
The gamma ray flux which corresponds
to this value of $\epsilon_{tot}$ is
$F_\gamma(E_\gamma>100MeV)\approx 1.2\cdot 10^{-7} cm^{-2} s^{-1}$,
which exceeds the EGRET limit on Coma by a factor $\sim 3$.

Thus, our conclusion can be stated as follows: if the radio and
X-ray non-thermal fluxes are due to synchrotron and ICS
of the same population of secondary electrons, then the SEM cannot
fit the two sets of observations withouth exceeding
the EGRET limit. In this case and within the assumptions used in this
paper, the SEM can be ruled out already on the basis of
the present data.

Is there any way to mitigate the strength of this result? We can
envision two possible avenues: from the point of view of data
analysis, it was already pointed out in \cite{fusco98} that
the rebinning of the hard X-ray data according with the OSSE energy
binning yields an hard X-ray flux which is a factor $\sim 2$ below 
the OSSE upper limit at $E \geq 40$ keV. 
This allows the magnetic field
to increase to values $B_\mu\approx 0.15-0.2$ and the corresponding
gamma ray fluxes to decrease slightly below the EGRET limit
for $\gamma\approx 2.1$, or slightly above it
for $\gamma=2.4$.

From the theoretical point of view, 
the ICS
origin of the hard X-ray excess observed in Coma cannot be taken
for granted  as pointed out in \cite{biermann}.
The previous authors propose, in fact, that the turbulence in the
ICM may be able to accelerate stochastically a fraction of
thermal electrons producing a non-thermal tail
of the electron distribution without appreciably
change the energy budget of the electrons in the IC gas.
Such non-thermal tail can easily reproduce
the SAX data through bremmstrahlung emission.
The main consequence of this is that the hard X-ray flux becomes
completely decoupled from the radio halo flux and no constraint
on the strength of the IC magnetic field can be derived.
Therefore, the large $B_{\mu}$ field solutions illustrated in
Figs.~1b and 1c
may well fit the radio spectrum and contribute only
very marginally to the hard X-ray spectrum.
In this case the gamma ray flux for Coma predicted by
the SEM is well below the EGRET limit.

As for the UV flux, the small discrepancy we found might be
due to the specific choice of the CR spectrum at low energy. 
In any case, the origin of the UV excess is still under  debate, 
and a thermal contribution cannot be excluded
(see discussion in \cite{ab98}).

We did not discuss here the additional issue
concerning the nature of the claimed steepening in
the radio halo spectrum at frequencies
$> 1.4$ GHz.
Whether this is a real feature or
an instrumental artifact is not yet clear (see discussion in
\cite{deiss}).
However, if future measurements will
confirm this result, the SEM will again have problems since
the electron ageing, often invoked to describe the steepening of the
spectrum, is not effective in the SEM because
secondary electrons are continuously produced
through CR interactions and replenish the radio halo spectrum,
even if the CR source is no longer active.
On the other hand, the SEM predicts a steepening of the
radio spectrum at high frequency, where CR begin to be not
confined by the IC magnetic field (see Section \ref{radio}
for details), but at present such
a steepening does not seem sufficient to reproduce the data
at $\nu\simgt 2.7$ GHz, assuming that these results
are not due to any instrumental bias.

\section{Discussion and conclusions}\label{discussion}

In this paper, we calculated the spectra of the non-thermal
radio, X-ray, UV and gamma ray emission
expected from clusters of galaxies in the
context of the secondary electron model, and we compared our
predictions to the available data for the Coma cluster.
The calculations have been carried out in the case of a
single source of CR and in the case of a homogeneous
injection of CR in the ICM.

For the case of a single CR source and adopting a Kolmogorov
spectrum of the magnetic fluctuations in the ICM, the
radio data between $30$ MHz and $1.4$ GHz
can be fitted in the context of the SEM for a wide range
of values of the parameters (specifically the
value of the magnetic field and the spectrum of the
injected CR). In particular, for magnetic fields of the
order of $1-2\mu G$ the CR luminosities needed to
fit the radio data are completely consistent with the
presence of ordinary sources of CR in clusters, as
considered in \cite{bbp} ($L_p\leq 10^{44}$ erg/s).

Moreover the model has the noticeable property
that electrons are produced {\it in situ} at any time,
so that there is no need for a reacceleration mechanism which is
necessary in the primary electron model in order to make it
possible for electrons to reach distances of $\sim 1$ Mpc (the scale
of the Coma radio halo) withouth appreciable energy losses.

Together with the diffuse radio emission, non-thermal X-rays
(due to ICS of secondary electrons) and gamma rays (mainly due to
neutral pion decay and to bremsstrahlung of secondary electrons) are 
produced.

The comparison of our predictions with the available X-ray and
gamma ray data allows us to put strong constraints
on the SEM as a plausible explanation for the
origin of the radio halos in Coma-like clusters.
In fact, if the X-ray emission is due to ICS of the same
electron population which is responsable for the diffuse
radio emission, then  low values
of the magnetic field ($B_\mu\sim 0.1$) and correspondingly
large CR luminosities ($L_p \simgt 5 \cdot 10^{45}$ erg/s)
are required, which in turn imply a flux of gamma rays
at $E_{\gamma} \geq 100$ MeV exceeding the EGRET limit
by a factor $\sim 2$ for a power index of the injection
spectrum of CR $\gamma=2.1$, and by a factor $\sim 7$ for
$\gamma=2.4$.\par

This result requires some more comments:
assuming a cluster temperature of $T=8.21$ keV, and a typical
IC gas density in the inner $\sim 1$ Mpc of the cluster of 
$n_H\sim 10^{-3} cm^{-3}$, 
the total thermal energy in the cluster can be estimated to be
$\epsilon_{th}\sim 2\cdot 10^{63}$ erg. Now, if we follow
\cite{lieu98} in requiring equipartition between thermal and non-
thermal energy in the ICM, this translates into a CR luminosity,
averaged over the age of the cluster, of  $L_p^{eq}\sim 5\cdot
10^{45}$ erg/s, which is of the same order of the CR 
luminosities required
in the SEM to reproduce the radio and hard X-ray emission
from Coma. 
However, as we have shown in Section 5, in this case the
corresponding gamma ray luminosities contributed by
pion decay can very easily 
exceed the EGRET limit or fall short of this limit by a very
small amount
This conclusion is not changed if a homogeneous
injection of CR in the ICM is assumed, as discussed in
Section \ref{coma}.

Though this is not enough to rule out the possibility of CR in
equipartition in clusters like Coma, it certainly represents a strong
constraint on this possibility. 
The next generation gamma ray satellites (INTEGRAL, GLAST)
will answer this question definitely. 
In fact, future gamma ray
observations will tell us not only the flux of gamma rays but also the
nature of the process responsible for that: 
specifically, if the bulk of gamma rays comes
from pion decay in the ICM, there is a unique signature, represented by a
bump in the gamma ray spectrum at $\sim 70$ MeV
(see Fig. 3).

Is there an alternative way of interpreting the combined radio and hard
X-ray data that does not require equipartition CR energy densities?
It was recently proposed in \cite{biermann} that the hard X-ray flux
from Coma can as well be interpreted as bremmstrahlung emission
from a supra-thermal electron tail developed
in the thermal electron distribution due to stochastic
acceleration in the turbulent ICM.
Clearly, these last electrons are not relevant for the
radio halo emission, so that a non-thermal population of
electrons in the $GeV$ energy range is required.
If this will turn out to be the explanation of the hard X-ray
excess of Coma, then the observed radio and non-thermal X-ray fluxes
are no longer strictly related.
As a consequence,  the radio halo emission can be produced by a
population of secondary electrons radiating in
a $B \sim 1\mu G$ average magnetic field,
without overproducing gamma rays (see Table 1) and giving a
marginal contribution to the non-thermal X-ray flux by  ICS.

In any case, it is important to realize that
the gamma ray limit, widely discussed in this paper, appears
to be relevant whatever the process
of production of the radiating electrons is. Therefore, the
EGRET upper limit must be considered as binding in the SEM, as well
as in other models for the non-thermal processes in clusters
of galaxies, whenever large CR energy densities are invoked.
This aspect was also recognized in \cite{lieu98} but the importance
of the gamma rays from pion decay was largely underestimated.

Finally, we want to discuss briefly the problem of the rarity of the
radio halos in clusters of galaxies.
This problem needs a definitive explanation in every model
for the origin of radio halos.
At present, only half a dozen clusters have been shown to have
prominent radio halos (see \cite{fg96} for a review) and it is
natural to look for the reason for such a rarity.
Most of the radio halo clusters have high temperatures
($T \simgt 7$ keV) and
X-ray luminosities $L_X \simgt {\rm a~few} ~ 10^{44}$ erg/s
and show evidences of recent merging
(Coma \cite{bh92}, A2319 \cite{fg97b}, A2163 \cite{mark}, A2218)
and/or of non--thermal pressure support of their IC medium
(A85 \cite{lm94})

There are a number of factors that can play a relevant role:
first of all, the CR injection in the ICM
depends on the morphology of the
cluster (e.g. on the number of active galaxies in a cluster)
and on the hystory of the cluster (e.g. on the occurrence of
recent mergings).
Thus, it is conceivable that only a few clusters have powerful CR
sources operating for long enough in the ICM to provide high powers in
accelerated particles.
Secondly, the different physical conditions present in the ICM
determine quite a variety of values for the magnetic field and for
the diffusion coefficient.
In fact, we have shown in this
paper that, at fixed $L_p$, changing the average IC magnetic field
from $1\mu G$ to $0.1\mu G$ decreases the radio halo flux by a factor
$\simgt 100$, so that very weak radio halos can fall
below the sensitivity threshold of the current experiments.

The origin of radio halos is, in our
opinion, strictly related to the mechanism of production of
the high energy electrons and to
the physical mechanisms which create
the environment in which they radiate, and can hardly
be envisioned at the present stage of observations.
The combination of the next generation, high-sensitivity radio, soft
and hard gamma ray experiments
will provide a zoo of non--thermal
phenomena in galaxy clusters and will
hopefully shed a new light on this problem.

\vskip 1.truecm
{\bf Aknowledgements}\par
We are grateful to an anonymous Referee for his comments and
suggestions which allowed us to improve considerably the presentation
of the paper.
The research of PB is funded by a INFN fellowship
at the University of Chicago.
S.C. also acknowledges useful discussions with L. Feretti, 
G. Giovannini and R. Fusco-Femiano.

\clearpage

{\Large
\begin{center}
\begin{table}
\caption{Summary of the fitting parameter values
\label{table1}}
\vskip 0.3truecm
\begin{tabular}{| c | c | c | c | c |}
\hline
\hline
$B_\mu$ & $\gamma$ & $\frac{L_p}{10^{44}erg s^{-1}}$ &
$\frac{F_\gamma(E_\gamma\geq 100MeV)}{F_\gamma^{EGRET}
(E_\gamma\geq 100MeV)}$ & $F_{\gamma}^{brem}/F_{\gamma}^{\pi^0}$
\\ \hline

$0.1$ & $2.1$ & $50$ & $1.93$ & 0.13\\
$0.1$ & $2.4$ & $180$ & $7.15$ & 0.10 \\
$1$ & $2.1$ & $0.35$ & $1.8\cdot 10^{-2}$ & 0.14 \\
$1$ & $2.4$ & $1$ & $4.5\cdot 10^{-2}$ & 0.11\\
$2$ & $2.1$ & $0.1$ & $5.3\cdot 10^{-3}$  & 0.12\\
$2$ & $2.4$ & $0.23$ & $1.1\cdot 10^{-2}$ & 0.095\\

\hline
\hline
\end{tabular}
\end{table}
\end{center}
}

\clearpage

\centerline{\bf Figure captions}
\vskip 1.truecm

{\bf Figure 1}~~
Spectrum of the diffuse radio emission from the Coma cluster for
different values of the average intracluster magnetic field:
$B=0.1~\mu G$ (left panel), $B=1~\mu G$ (central panel) and
$B=2~\mu G$ (right panel).
A King density profile has been used with
$\beta_{IC}=0.75, n_H^0=2.89 \cdot 10^{-3} ~cm^{-3}$ and $r_c=0.42$ Mpc
(we use here $h=1/2$).
For each panel the cases $\gamma=2.1$ (continuous curves) and
$\gamma=2.4$ (dashed curves) are shown.
The parameters that best fit the data in each figure are
listed in Table \ref{table1} and are obtained fitting the
radio halo data at $\nu\simgt 30$ MHz.
Data points are taken from \cite{gfer}.

{\bf Figure 2}~~
Spectrum  of the diffuse X-ray emission from Coma.
The three panels refer to the same values of the IC magnetic field as
in Fig. 1.
The shaded area shows the best fit to the HEAO1-A4 and GINGA
thermal emission data (open triangles) at $T=8.21 \pm 0.20$ keV
\cite{hughes}.
The OSSE upper limits \cite{osse} are indicated by the open circles.
The SAX data \cite{fusco98} are indicated by filled squares.
Arrows and labels show, for each panel, the energy ranges in which the
three different data sets are located.
Predictions of the SEM for $\gamma=2.1$ (continuous curves) and
$\gamma=2.4$ (dashed curves) are shown in each panel.

{\bf Figure 3}~~
The predicted differential gamma ray spectra from Coma, in the SEM. 
The three panels refer
to the values of the magnetic field indicated. The thick lines represent
the contribution of neutral pion decay, while the thin lines represent
the bremsstrahlung contribution of secondary electrons. Solid and
dashed curves are for $\gamma=2.1$ and $\gamma=2.4$ respectively.

\clearpage
\begin{figure}[thb]
 \begin{center}
  \mbox{\epsfig{file=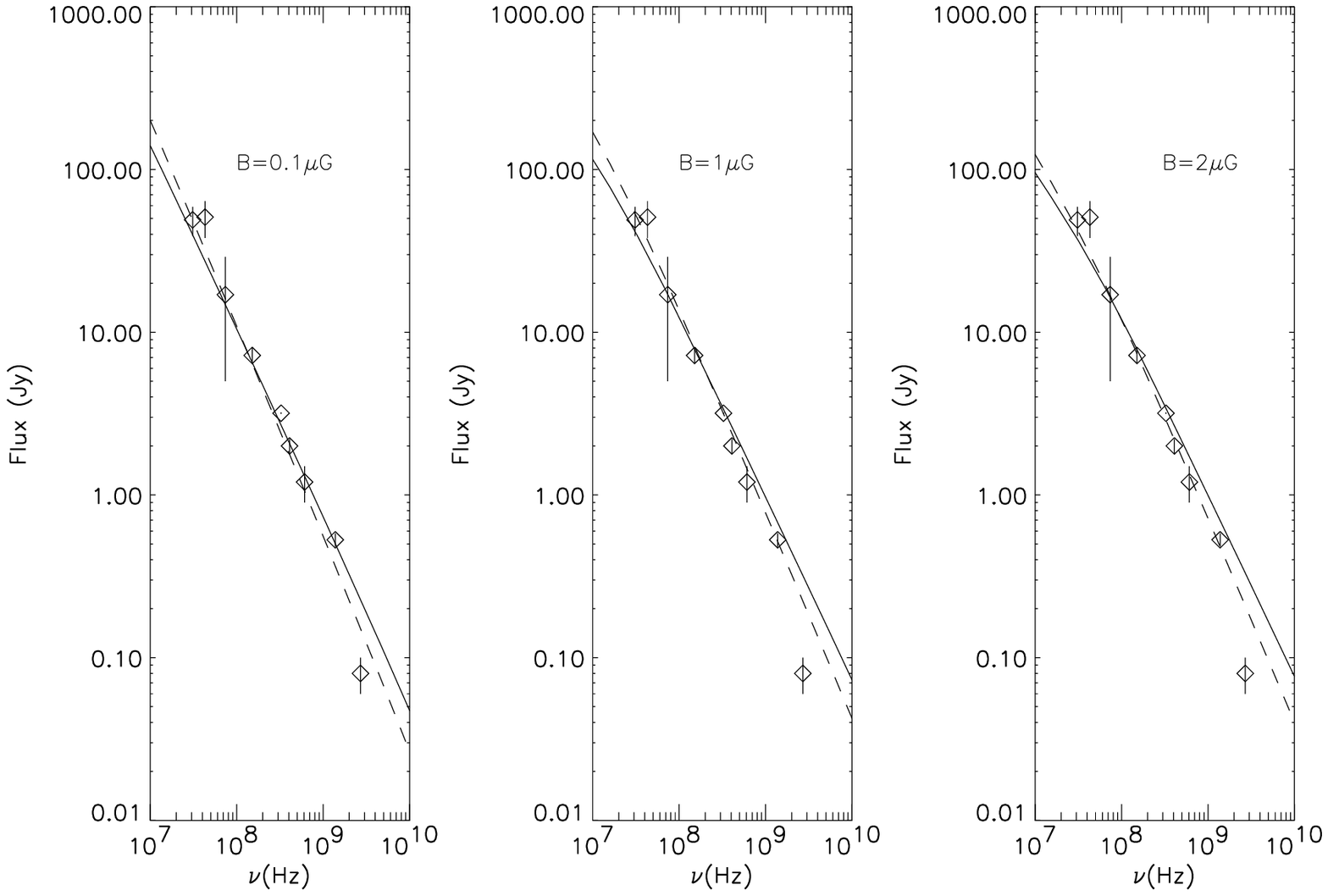,width=17.cm}}
  \caption{\em {
}}
 \end{center}
\end{figure}
\clearpage
\begin{figure}[thb]
 \begin{center}
  \mbox{\epsfig{file=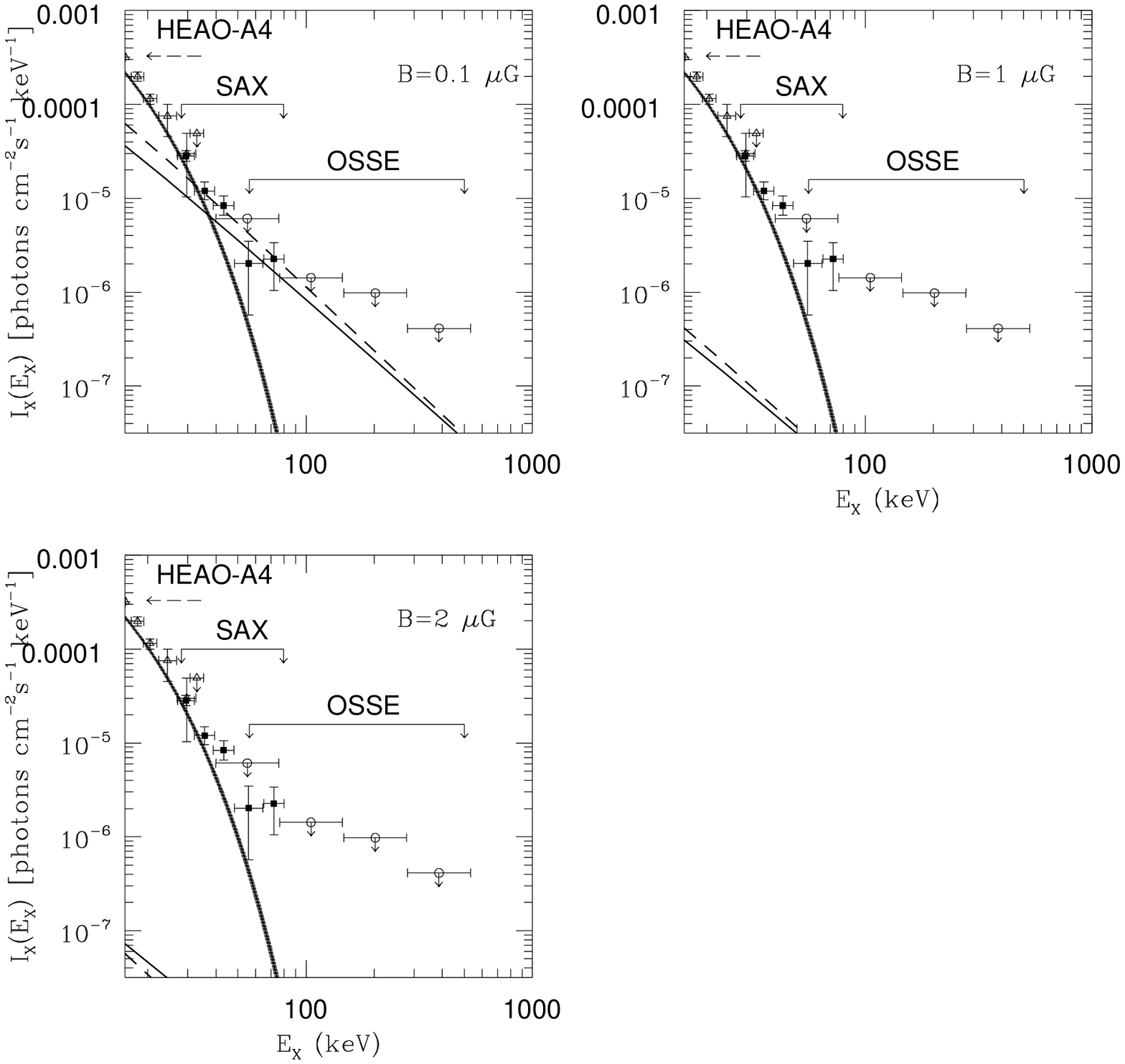,width=18.cm}}
  \caption{\em {
}}
 \end{center}
\end{figure}
\clearpage
\begin{figure}[thb]
 \begin{center}
  \mbox{\epsfig{file=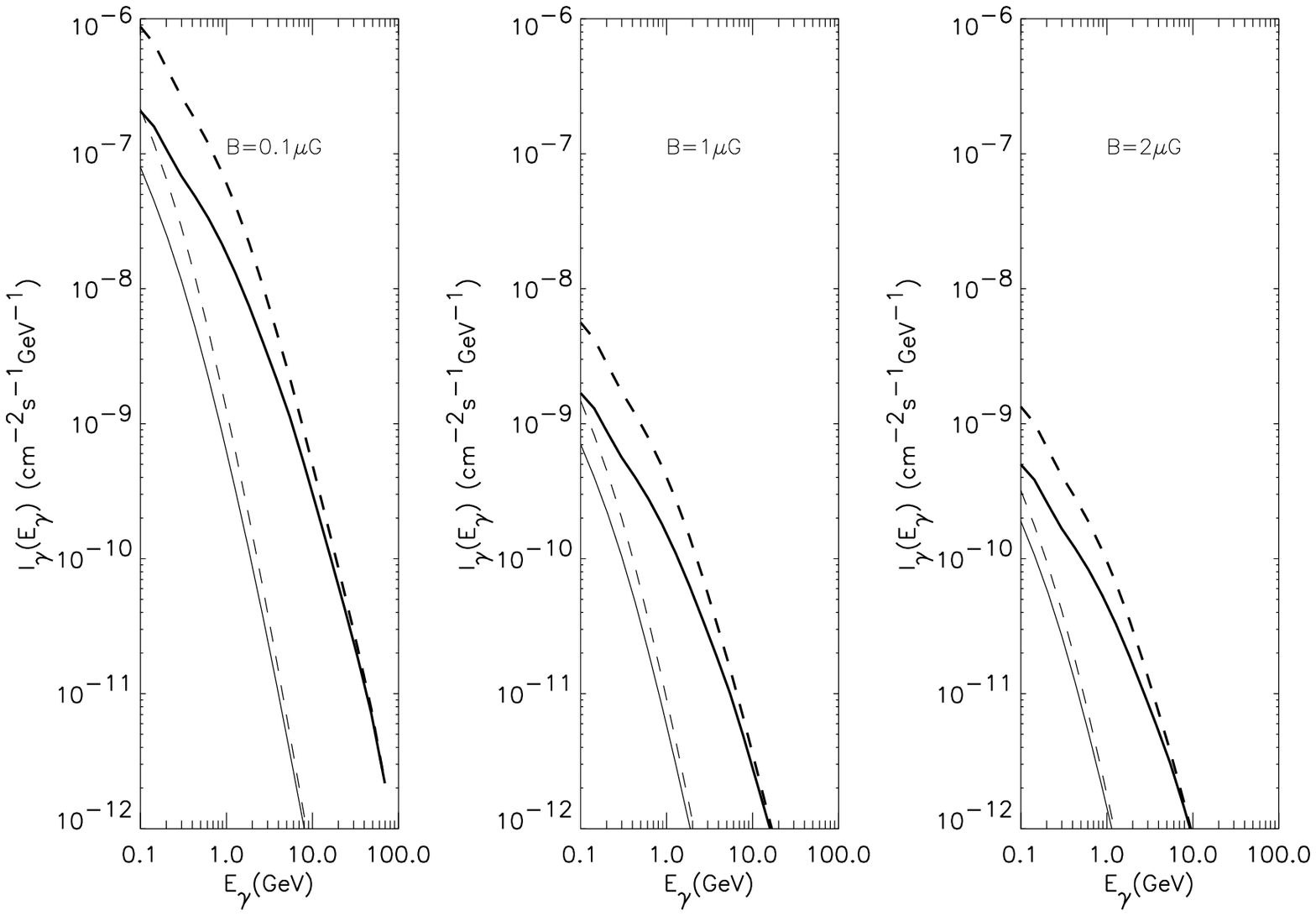,width=17.cm}}
  \caption{\em {
}}
 \end{center}
\end{figure}

\end{document}